\documentclass{article}
\usepackage{amssymb}

\input{tcilatex}

\begin{document}

\pagestyle{myheadings} \markright{\it 276-2} \vskip.5in

\begin{center}
%
%
\vskip.4in \textbf{Can two or more gauge bosons propagate in the light-front?%
}{\Large \textbf{\ }} \vskip.3in 
%
%
%
A.T.Suzuki\footnote{%
Email: \texttt{suzuki@ift.unesp.br }}\\[0pt]
Instituto de F\'{i}sica Te\'{o}rica-UNESP, 01405-900 S\~{a}o Paulo, Brazil. 
%
%
\\[0pt]
J.H.O.Sales\footnote{%
Email: \texttt{henrique@fatecsp.br}}\\[0pt]
Faculdade de Tecnologia de S\~{a}o Paulo-DEG, P\c{c}a. Coronel Fernando
Prestes, 01124-060 S\~{a}o Paulo-SP.
\end{center}

%
Gauge fields in the light-front are usually fixed via the $n\cdot A=0$
condition yielding the non-local singularities of the type $(k\cdot
n)^{-\alpha }=0$ and $\alpha =1,2,..$ in the gauge boson propagator which
must be addressed conveniently. In calculating this propagator for $n$
noncovariant gauge bosons those non-local terms demand the use of a
prescription to ensure causality. We show that from 2 gauge bosons onward
the implementation of such a prescription does not remove certain
pathologies such as the non existence of two or more free propagating gauge
bosons in the light-front form.\vskip.2in

%
%
%
%
%
%
%
%
%
%
%

\section{Light-front}

In 1949 Dirac \cite{dirac}, showed that it is possible to construct
dynamical forms from the description of a initial state of a given
relativistic system in any space-time surface whose lengths between points
have no causal connection. The dynamical evolution corresponds to the system
following a trajectory throught the hyper-surfaces. For example, the
hyper-surface $t=0$ is our three-dimensional space. It is invariant under
translations and rotations. However, in any transformation of inertial
reference frames which involves ``boosts'', the temporal coordinate is
modified, and therefore the hyper-surface in $t=0$. Other hyper-surfaces can
be invariant under some type of ``boost''. It is the case of the hyperplane
called null plane, defined by $x^+=t+z$, which in analogy to the usual
coordinate system, is commonly referred to as the ``time'' coordinate for
the front form (light front). For example, a ``boost'' in the $z$ direction
does not modify the null plane.

A point in the usual four-dimensional space-time is defined through the set
of coordinates $\left( x^{0},x^{1},x^{2},x^{3}\right)$, where $x^{0}$ is the
time coordinate, that is, $x^{0}=t$, with the usual convention of taking the
speed of light equal to unity, $c=1$. The other coordinates are the
three-dimensional Euclidean space coordinates $x^{1}=x$, $x^{2}=y$ and $%
x^{3}=z$.

The light-front coordinates are defined in terms of these by the following
relations: 
\begin{eqnarray}
x^{+}&=&x^{0}+x^{3},  \nonumber \\
x^{-}&=&x^{0}-x^{3},  \nonumber \\
\vec{x}^{\perp }&=& x^{1}\vec{i}+x^{2}\vec{j},
\end{eqnarray}
where $\vec{i}$ and $\vec{j}$ are the unit vectors in the direction of the
coordinates $x$ and $y$. The null plane is defined by $x^{+}=0$, that is,
this condition defines the hyper-surface which is tangent to the light cone,
the reason why some authors call those light-cone coordinates.

Note that for the usual four-dimensional Minkowski space-time whose metric $%
g^{\mu\nu}$ is defined such that its signature is $(1,-1,-1,-1)$ we have 
\begin{eqnarray}
x^{+}&=&x^{0}+x^{3}\;\;=\;\;x_0-x_3\;\;\equiv \;\;x_{-},  \nonumber \\
x^{-}&=&x^{0}-x^{3}\;\;=\;\;x_0+x_3\;\;\equiv \;\;x_{+},  \nonumber \\
\vec{x}^{\perp }&=& x^{1}\vec{i}+x^{2}\vec{j}\;\;=\;\;-x_{1}\vec{i}-x_2\vec{j%
}\;\;\equiv \;\;-x_{\perp},
\end{eqnarray}

The initial boundary conditions for the dynamics in the light front are
defined in this hyper-plane. Note that the axis $x^{+}$ is orthogonal to the
plane $x^{+}=0$. Therefore, a displacement of this hyper-surface for $%
x^{+}>0 $ is analogous to the displacement of the plane $t=0$ for $t>0$ of
the usual four-dimensional space-time. With this analogy we identify $x^{+}$
as the ``time'' coordinate for the null plane. Of course, since there is a
conspicuous discrete symmetry between $x^+ \leftrightarrow x^-$, one could
choose $x^-$ as his ``time'' coordinate. However, once chosen, one has to
stick to the convention adopted. We shall adhere to the former one.

The canonically conjugate momenta for the coordinates $x^{+},x^{-}$ and $%
x^{\perp }$ are defined respectively by: 
\begin{eqnarray}
k^{+}&=&k^{0}+k^{3},  \nonumber \\
k^{-}&=&k^{0}-k^{3},  \nonumber \\
k^{\perp }&=&\left( k^{1},k^{2}\right) .  \label{1.4}
\end{eqnarray}

The scalar product in the light front coordinates becomes therefore 
\begin{equation}
a^{\mu }b_{\mu }=\frac{1}{2}\left( a^{+}b^{-}+a^{-}b^{+}\right) - \vec{a}%
^{\perp }\cdot \vec{b}^{\perp },  \label{1.7}
\end{equation}
where $\vec{a}^{\perp }$ and $\vec{b}^{\perp }$ are the transverse
components of the four vectors. All four vectors, tensors and other entities
bearing space-time indices such as Dirac matrices $\gamma ^{\mu }$ can be
expressed in this new way, using components $(+,-,\perp)$.

From (\ref{1.7}) we can get the scalar product $x^{\mu}k_{\mu }$ in the
light front coordinates as: 
\begin{equation}
x^{\mu }k_{\mu }=\frac{1}{2}\left( x^{+}k^{-}+x^{-}k^{+}\right) - \vec{x}%
^{\perp }\cdot \vec{k}^{\perp }.  \label{1.8}
\end{equation}

Here again, in analogy to the usual four-dimensional Minkowski space-time
where such a scalar product is 
\begin{equation}
x^{\mu} k_{\mu}=x^0 k^0-\mathbf{{x}\cdot {k}}
\end{equation}
where $\mathbf{x}$ is the three-dimensional vector, with the energy $k^0$
associated to the time coordinate $x^0$, we have the light-front ``energy'' $%
k^-$ associated to the light-front ``time'' $x^+$. Note, however, that there
is a crucial difference between the two formulations: while the usual
four-dimensional space-time is Minkowskian, the light-front coordinates
projects this onto two sectorized Euclidean spaces, namely $(+,-)$, and $%
(\perp,\perp)$.

In the Minkowski space described by the usual space-time coordinates we have
the relation between the rest mass and the energy for the free particle
given by $k^{\mu }k_{\mu }=m^{2}$. Using (\ref{1.7}), we have 
\[
k^{\mu }k_{\mu }=\frac{1}{2}\left( k^{+}k^{-}+k^{-}k^{+}\right) - \vec{k}%
^{\perp }\cdot \vec{k}^{\perp }, 
\]
so that 
\begin{equation}
k^{-}=\frac{\overrightarrow{k}_{\perp }^{2}+m^{2}}{k^{+}}.  \label{1.10}
\end{equation}

Note that the energy of a free particle is given by $k^{0}=\pm \sqrt{m^{2}+%
\mathbf{{k}^{2}}}$, which shows us a quadratic dependence of $k^{0}$ with
respect to $\mathbf{k}$. These positive/negative energy possibilities for
such a relation were the source of much difficulty in the interpretation of
the negative energy particle states in the beginning of the quantum field
theory descriprion for particles, finally solved by the antiparticle
interpretation given by Feynman. In contrast to this, we have a linear
dependence between $(k^{+})^{-1}$ and $k^{-}$ (see Eq.(\ref{1.10})), which
immediately reminds us of the non-relativistic quantum mechanical type of
relationship for one particle state systems.

\section{Classical Propagator}

In a recent work of ours \cite{reex1}, we showed that a single Lagrange
multiplier defined by $(n\cdot A)(\partial \cdot A)$ with $n\cdot A=\partial
\cdot A=0$ at the classical level leads to a propagator in the light-front
gauge that has no residual gauge freedom left.

Thus, for the relevant gauge fixing term that enters in the Lagrangian
density which we define as 
\begin{equation}
(n\cdot A)(\partial \cdot A)=0,
\end{equation}
gives for the Abelian gauge field Lagrangian density: 
\begin{equation}
\mathcal{L}=-\frac{1}{4}F_{\mu \nu }F^{\mu \nu }-\frac{1}{2\alpha }\left(
2n_{\mu }A^{\mu }\partial _{\nu }A^{\nu }\right) =\mathcal{L}_{\mathrm{E}}+%
\mathcal{L}_{GF}  \label{8}
\end{equation}
where the gauge fixing term is conveniently written so as to symmetrize the
indices $\mu $ and $\nu $, and the gauge parameter can assume complex
values. By partial integration and considering that terms which bear a total
derivative don't contribute and that surface terms vanish since $%
\lim\limits_{x\rightarrow \infty }A^{\mu }(x)=0$, we have 
\begin{equation}
\mathcal{L}_{\mathrm{E}}=\frac{1}{2}A^{\mu }\left( \square g_{\mu \nu
}-\partial _{\mu }\partial _{\nu }\right) A^{\nu }  \label{9}
\end{equation}
and 
\begin{equation}
\mathcal{L}_{GF}=-\frac{1}{\alpha }(n\cdot A)(\partial \cdot A)=-\frac{1}{%
2\alpha }A^{\mu }\left( n_{\mu }\partial _{\nu }+n_{\nu }\partial
_{\mu}\right) A^{\nu }  \label{25}
\end{equation}
so that 
\begin{equation}
\mathcal{L}=\frac{1}{2}A^{\mu }\left( \square g_{\mu \nu }-\partial _{\mu
}\partial _{\nu }-\frac{1}{\alpha }(n_{\mu }\partial _{\nu }+n_{\nu
}\partial _{\mu })\right) A^{\nu }  \label{11}
\end{equation}

To find the gauge field propagator we need to find the inverse of the
operator between parenthesis in (\ref{11}). That differential operator in
momentum space is given by $O_{\mu \nu }(k)=-k^{2}g_{\mu \nu }+k_{\mu
}k_{\nu }+\frac{1}{\alpha }\left( n_{\mu }k_{\nu }+n_{\nu }k_{\mu }\right) $%
, so that the propagator of the field, which we call $S^{\mu \nu }(k)$, must
satisfy the following equation $O_{\mu \nu }S^{\nu \lambda }\left( k\right)
=\delta _{\mu }^{\lambda }$, where $S^{\nu \lambda }(k)$ can now be
constructed from the most general tensor structure that can be defined,
i.e., all the possible linear combinations of the tensor elements that
composes it (the most general form includes the light-like vector $m_{\mu }$
dual to the $n_{\mu }$ \cite{progress} -- but for our present purpose it is
in fact indifferent): 
\begin{eqnarray}
G^{\mu \nu }(k) &=&g^{\mu \nu }A+k^{\mu }k^{\nu }B+k^{\mu }n^{\nu }C+n^{\mu
}k^{\nu }D+k^{\mu }m^{\nu }E+  \nonumber \\
&&+m^{\mu }k^{\nu }F+n^{\mu }n^{\nu }G+m^{\mu }m^{\nu }H+n^{\mu }m^{\nu
}I+m^{\mu }n^{\nu }J  \label{a2}
\end{eqnarray}

Then, it is a matter of straightforward algebraic manipulation to get the
relevant propagator in the light-front gauge, namely, 
\begin{equation}  \label{classicprop}
S^{\mu \nu }(k)=-\frac{1}{k^{2}}\left\{ g^{\mu \nu }-\frac{k^{\mu }n^{\nu
}+n^{\mu }k^{\nu }}{k^{+}}+\frac{n^{\mu }n^{\nu }}{(k^{+})^{2}}k^{2}\right\}
\,  \nonumber
\end{equation}

\section{Quantum gauge boson propagator}

The Feynman quantum propagator for the gauge boson can be derived
integrating over all the momenta in (\ref{classicprop}). Projecting out this
propagator on to the light-front we get a gauge boson particle propagating
at equal light-front times. We are going to restrict our calculation to the
total momentum $P^{+}$ positive and corresponding forward light-front time
propagation. In this case the propagator from $x^{+}=0$ to $x^{+}>0$ is
given by: 
\begin{equation}
\widetilde{S}^{(1)\mu \nu }(x_{1}^{\mu })=i\;\int \frac{d^{4}k_{1}}{\left(
2\pi \right) ^{4}}\frac{N^{\mu \nu }e^{-ik_{1}^{\mu }x_{1\mu }}}{%
k_{1}^{2}+i\varepsilon }.  \label{pc1}
\end{equation}
where 
\begin{equation}
N^{\mu \nu }=\frac{-g^{\mu \nu }k_{1}^{+2}+\left( k_{1}^{\mu }n^{\nu
}+n^{\mu }k_{1}^{\nu }\right) k_{1}^{+}-n^{\mu }n^{\nu }k_{1}^{2}}{k_{1}^{+2}%
}.  \label{n}
\end{equation}

Note that because of the structure of the light-front propagator (\ref
{classicprop}) only three of the component projections are non vanishing,
namely, 
\begin{equation}
N^{\perp \perp }=-g^{\perp \perp },\qquad N^{\perp -}=\frac{
n^{-}k_{1}^{\perp }}{k_{1}^{+}}\qquad N^{--}=\frac{n^{-}n^{-}k_{1}^{\perp 2}%
}{k_{1}^{+2}}  \label{cn}
\end{equation}
At equal light-front times, we have: 
\begin{equation}
\widetilde{S}^{(1)\mu \nu }(x^{+})=\frac{i}{2}\int \frac{dk_{1}^{-}}{\left(
2\pi \right) }\frac{N^{\mu \nu }e^{-\frac{i}{2}k_{1}^{-}x^{+}}}{%
k_{1}^{+}\left( k_{1}^{-}-\frac{(k_{1}^{\perp})^{2}}{k_{1}^{+}}+\frac{%
i\varepsilon }{k_{1}^{+}}\right) }.  \label{1b2}
\end{equation}
so that, in terms of the component projections we have immediately 
\begin{eqnarray}
\widetilde{S}^{++}\:\:=\:\: \widetilde{S}^{+-} \:\:=\:\: \widetilde{S}%
^{+\perp } & = & 0.  \label{cs} \\
\widetilde{S}^{\perp \perp }\:\: \neq \:\: \widetilde{S}^{\perp -}\:\:
\neq\:\: \widetilde{S}^{--} & \neq & 0  \nonumber
\end{eqnarray}

The $(\perp,\perp)$ component presents no particular difficulty in
evaluation nor does it present any overwhelming troublesome feature.
However, the components $(\perp,-)$ and $(-,-)$, while having similar
technical difficulties in the performing of the relevant computation, do
come with an overwhelming troublesome result which will become clearer as we
proceed further on in our analysis. This feature is nothing more nothing
less than the troublesome zero mode problem in the light front. We shall
therefore restrict ourselves to the analysis of the $(\perp,-)$ component.

\noindent In terms of Fourier transform we have: 
\begin{equation}
S^{(1)\mu \nu }(p^{-})=\int dx^{+}e^{\frac{i}{2}p^{-}x^{+}}\widetilde{S}%
^{(1)\mu \nu }(x^{+})\ ,  \label{tf}
\end{equation}
so that the component $S^{\perp -}$ is: 
\begin{eqnarray}
S^{(1)\perp -}(p^{-}) &=&i\int dk_{1}^{-}\frac{k_{1}^{\perp }\;n^{-}\delta
\left( p^{-}-k_{1}^{-}\right) }{\left( k_{1}^{-}-\frac{k_{1\perp }^{2}}{%
k_{1}^{+}}+\frac{i\varepsilon }{k_{1}^{+}}\right) }\left[ \frac{1}{\left(
k_{1}^{+}\right) ^{2}}\right] _{\mathrm{ML}}  \nonumber \\
&=&i\int dk_{1}^{-}\frac{k_{1}^{\perp }\;n^{-}\delta \left(
p^{-}-k_{1}^{-}\right) }{\left( k_{1}^{-}-\frac{k_{1\perp }^{2}}{k_{1}^{+}}+%
\frac{i\varepsilon }{k_{1}^{+}}\right) }  \nonumber \\
&&\times \left[ \frac{k_{1}^{-}}{k_{1}^{+}\left( k_{1}^{-}+\frac{%
i\varepsilon }{k_{1}^{+}}\right) }\right] _{\mathrm{ML}}^{2}  \label{prlf1}
\end{eqnarray}
where 
\[
\delta \left( \frac{p^{-}-k_{1}^{-}}{2}\right) =\frac{1}{2\pi }\int dx^{+}e^{%
\frac{i}{2}\left( p^{-}-k_{1}^{-}\right) x^{+}} 
\]
and the index \textrm{ML} stands for the Mandelstam-Leibbrandt prescription 
\cite{ml} for the treatment of the $(k^{+})^{-1}$ poles, namely, 
\begin{equation}
\left[ \frac{1}{k^{+}}\right] _{\mathrm{ML}}=\lim_{\varepsilon \rightarrow 0}%
\left[ \frac{k^{-}}{k^{+}k^{-}+i\varepsilon }\right] _{\mathrm{ML}}
\label{ml}
\end{equation}

\noindent The result is: 
\begin{equation}
S^{(1)\perp -}(p^{-})=\frac{\theta (p^{+})\;p^{\perp }n^{-}}{p^{+2}}\frac{i}{%
\left( p^{-}-K_{0}^{(1)-}+i\varepsilon \right) },  \label{1b3}
\end{equation}
where we have introduced the definition 
\begin{equation}
K_{0}^{(1)-}=\frac{p_{\perp }^{2}}{p^{+}},  \label{1b4}
\end{equation}
as the light-front Hamiltonian of the free one-particle system. Note that
for $x^{+}<0$, the $S^{(1)}(x^{+})=0$ because $p^{+}>0$. Moreover, observe
that $S^{(1)\perp -}(p^{-})$ is in an operator form with respect to $p^{+}$
and $\vec{p}_{\perp }$. Consequently we have a clear manifestation of the
zero mode problem in the factor $(p^+)^{-2}=0$.

\section{\noindent Two gauge boson propagator}

The two-boson gauge propagator can be derived from the covariant propagator
for two particles propagating at equal light-front times. Without loosing
generality, we are going to restrict our calculation to the total momentum $%
P^{+}$ positive and corresponding forward light-front time propagation. In
this case the propagator from $x^{+}=0$ to $x^{+}>0$ is given by: 
\begin{eqnarray}
\widetilde{S}^{(2)\mu \nu ;\alpha \beta }(x^{\prime \mu }{},x^{\mu })
&=&\int \frac{d^{4}k_{1}}{\left( 2\pi \right) ^{4}}\frac{d^{4}k_{2}}{\left(
2\pi \right) ^{4}}\frac{iN^{\mu \nu }e^{-ik_{1}^{\mu }\left( x_{1\mu
}^{\prime }-x_{1\mu }\right) }}{k_{1}^{2}+i\varepsilon }  \nonumber \\
&&\frac{iN^{\alpha \beta }e^{-ik_{2}^{\mu }\left( x_{2\mu }^{\prime
}-x_{2\mu }\right) }}{k_{2}^{2}+i\varepsilon }.  \nonumber
\end{eqnarray}

At equal light-front times $x_{1}^{+}=x_{2}^{+}=0$ and $x^{\prime
+}_1=x^{\prime +}_2=x^{+}$, the propagator is written as: 
\begin{equation}
\widetilde{S}^{(2)}(x^{+})=\widetilde{S}_{1}^{(1)}(x^{+})\widetilde{S}
_{2}^{(1)}(x^{+}),  \label{s2a}
\end{equation}
where the one-body propagators, $\widetilde{S}_{i}^{(1)}$, corresponding to
the light-front propagators of particles $i=1$ or $2$, are defined by Eq.(%
\ref{pc1}). We have explicitly: 
\begin{eqnarray}
\widetilde{S}^{(2)\mu \nu ;\alpha \beta }(x^{+}) &=&-\frac{2}{4}\int \frac{
dk_{1}^{-}}{\left( 2\pi \right) }\frac{dk_{2}^{-}}{\left( 2\pi \right) } 
\frac{N^{\mu \nu }e^{-\frac{i}{2}k_{1}^{-}x^{+}}}{k_{1}^{+}\left( k_{1}^{-}- 
\frac{k_{1}^{\perp 2}-i\varepsilon }{k_{1}^{+}}\right) }  \nonumber \\
&\times &\frac{N^{\alpha \beta }e^{-\frac{i}{2}k_{2}^{-}x^{+}}}{
k_{2}^{+}\left( k_{2}^{-}-\frac{k_{2}^{\perp 2}-i\varepsilon }{k_{2}^{+}}
\right) }.
\end{eqnarray}

The Fourier transform to the total light-front energy $P^{-}$ is given by 
\begin{equation}
S^{(2)\mu \nu ;\alpha \beta }(P^{-})=\frac{1}{2}\int dx^{+}e^{\frac{i}{2}
P^{-}x^{+}}\widetilde{S}^{(2)\mu \nu ;\alpha \beta }(x^{+})\ .
\end{equation}

As before, we can recognize immediately that (we ommit the $(2)$ index as
well as the $P^{-}$ dependence for shortness) 
\begin{eqnarray}
S^{++,++} &=&S^{+-,++}::=::S^{++,--}::=::S^{++,+\perp }::=::S^{++,\perp
\perp }::=::0  \nonumber \\
S^{\perp -,--} &\neq &S^{\perp -,\perp \perp }::\neq ::S^{--,\perp \perp
}::\neq ::S^{--,--}::\neq ::S^{\perp -,\perp -}::\neq ::S^{\perp \perp
,\perp \perp }\neq 0  \nonumber \\
&&
\end{eqnarray}
which result in 
\begin{eqnarray}
S^{(2)\perp -,\perp -}(P^{-}) &=&-\frac{1}{\left( 2\pi \right) }\int \frac{%
dk_{1}^{-}}{k_{1}^{+}\left( P^{+}-k_{1}^{+}\right) }  \label{plf2} \\
&\times &\left[ \frac{k_{1}^{-}-k_{1\mathrm{on}}^{-}}{k_{1}^{+}\left(
k_{1}^{-}-k_{1\mathrm{on}}^{-}+\frac{i\varepsilon }{k_{1}^{+}}\right) }%
\right] _{\mathrm{ML}}^{2}\frac{N_{1}^{\perp -}}{\left( k_{1}^{-}-\frac{%
k_{1\perp }^{2}}{k_{1}^{+}}+\frac{i\varepsilon }{k_{1}^{+}}\right) } 
\nonumber \\
&\times &\left[ \frac{P^{-}-k_{1}^{-}-k_{2\mathrm{on}}^{-}}{\left(
P^{+}-k_{1}^{+}\right) \left( P^{-}-k_{1}^{-}-k_{2\mathrm{on}}^{-}+\frac{%
i\varepsilon }{P^{+}-k_{1}^{+}}\right) }\right] _{\mathrm{ML}}^{2}  \nonumber
\\
&\times &\frac{N_{2}^{\perp -}}{\left( P^{-}-k_{1}^{-}-\frac{\left( P_{\perp
}-k_{1}\right) ^{2}}{k_{1}^{+}}+\frac{i\varepsilon }{P^{+}-k_{1}^{+}}\right) 
}\ ,  \nonumber
\end{eqnarray}
where $P^{-,+,\perp }=k_{1}^{-,+,\perp }+k_{2}^{-,+,\perp }$.

We perform the analytical integration in the $k_{1}^{-}$ momentum by
evaluating the residue at the pole 
\begin{eqnarray*}
k_{1}^{-} &=&k_{1\mathrm{on}}^{-}-\frac{i\varepsilon }{k_{1}^{+}}; \\
k_{1}^{-} &=&P^{-}-k_{2\mathrm{on}}^{-}+\frac{i\varepsilon }{P^{+}-k_{1}^{+}}%
.
\end{eqnarray*}
It implies that only $k_{1}^{+}$ in the interval $0<k_{1}^{+}<K^{+}$ gives a
nonvanishing contribution to the integration. It implies that only $%
k_{1}^{+} $ in the interval $0<k_{1}^{+}<P^{+}$ gives a nonvanishing
contribution to the integration. The result is 
\begin{equation}  \label{perpm}
S^{(2)\perp -,\perp -}(P^{-})=\frac{\theta (k_{1}^{+})\theta
(P^{+}-k_{1}^{+})(k_{1}^{\perp }n^{-})(P^{\perp }-k_{1}^{\perp })n^{-}}{%
k_{1}^{+2}\left( P^{+}-k_{1}^{+}\right) ^{2}}\frac{i}{\left(
P^{-}-K_{0}^{(2)-}+i\varepsilon \right) },  \label{pm}
\end{equation}
and for the $(\perp \perp ,\perp \perp )$ we have 
\begin{equation}
S^{(2)\perp \perp ,\perp \perp }(P^{-})=\frac{\theta (k_{1}^{+})\theta
(P^{+}-k_{1}^{+})}{k_{1}^{+}\left( P^{+}-k_{1}^{+}\right) }\frac{i\left(
-g^{\perp \perp }\right) \left( -g^{\perp \perp }\right) }{\left(
P^{-}-K_{0}^{(2)-}+i\varepsilon \right) },  \label{pp}
\end{equation}
where 
\begin{equation}
K_{0}^{(2)-}=\frac{k_{1\mathrm{on}}^{2}}{k_{1}^{+}}+\frac{k_{2\mathrm{on}%
}^{2}}{P^{+}-k_{1}^{+}},  \label{k02}
\end{equation}
$K_{0}^{(2)-}$ is the light-front Hamiltonian of the free two-particle
system. For $x^{+}\ <\ 0$, $S^{(2)}(x^{+})\ =\ 0$ due to our choice of $%
P^{+}>0$. Observe that $S^{(2)}(P^{-})$ is written in Eq.(\ref{pm}) and Eq.(%
\ref{pp}) in operator form with respect to $k^{+}$ and $\vec{k}_{\perp }$.
Again we have problems of a divergent factor for $k_{1}^{+}=0$ in (\ref{pm}).

\section{Conclusion}

Projecting the Feynman covariant space propagator in light-front coordinates
and using the Mandelstam-Leibbrandt prescritpion to treat $k^-=0$
singularities we get propagation of one or two bodies in the light-front for
some components such as $(\perp -, \perp -)$ and $(\perp \perp, \perp \perp)$%
. Now, the $(\perp -, \perp -)$ component presents the zero mode problem for 
$k^+=0$. The same happens with other nonvanishing components, except for the 
$(\perp \perp, \perp \perp)$ component where there is no singularity of this
type.

We observe that even with the use of Leibbrandt-Mandelstam prescription, it
was not possible to remove the built-in singularity in $k^+=0$.

\vspace{.5cm}

\textbf{Acknowledgments:} A.T.S thanks partial support from CNPq
(Bras\'{\i}lia, DF). J.H.O.Sales is supported by FAPESP (S\~{a}o Paulo, SP).

\end{document}